\documentclass[sigconf]{acmart}
\usepackage{tipa}
\usepackage{enumitem}
\usepackage{graphicx} 
\usepackage{listings}

\AtBeginDocument{%
  }

\copyrightyear{2026}
\acmYear{2026}
\setcopyright{cc}
\setcctype{by}
\acmConference[CHI '26]{Proceedings of the 2026 CHI Conference on Human Factors in Computing Systems}{April 13--17, 2026}{Barcelona, Spain}
\acmBooktitle{Proceedings of the 2026 CHI Conference on Human Factors in Computing Systems (CHI '26), April 13--17, 2026, Barcelona, Spain}
\acmPrice{}
\acmDOI{10.1145/3772318.3791266}
\acmISBN{979-8-4007-2278-3/2026/04}

\begin{document}
\title{Voice-Based Chatbots for English Speaking Practice in Multilingual Low-Resource Indian Schools: A Multi-Stakeholder Study }

\author{Sneha Shashidhara}
\affiliation{%
  \institution{Centre for Social and Behaviour Change, Ashoka University}
  \city{New Delhi, New Delhi}
  \country{India}}
\email{sneha.shashidhara@ashoka.edu.in}

\author{Vivienne Bihe Chi}
\affiliation{%
  \institution{University of Pennsylvania}
  \city{Philadelphia, Pennsylvania}
  \country{USA}}
\email{vchi@seas.upenn.edu}

\author{Abhay P Singh}
\affiliation{%
  \institution{University of Pennsylvania}
  \city{Philadelphia, Pennsylvania}
  \country{USA}}
\email{sabhay@seas.upenn.edu}

\author{Lyle Ungar}
\affiliation{%
\institution{University of Pennsylvania}
  \city{Philadelphia, Pennsylvania}
  \country{USA}}
\email{ungar@cis.upenn.edu}

\author{Sharath Chandra Guntuku}
\affiliation{%
\institution{University of Pennsylvania}
  \city{Philadelphia, Pennsylvania}
  \country{USA}}
\email{sharathg@seas.upenn.edu}

 \renewcommand{\shortauthors}{Shashidhara et al.}

\begin{abstract}
Spoken English proficiency is a powerful driver of economic mobility for low-income Indian youth, yet opportunities for spoken practice remain scarce in schools. We investigate the deployment of a voice-based chatbot for English conversation practice across four low-resource schools in Delhi. Through a six-day field study combining observations and interviews,  we captured the perspectives of students, teachers, and principals. Findings confirm high demand across all groups, with notable gains in student speaking confidence. Our multi-stakeholder analysis surfaced a tension in long-term adoption vision: students favored open-ended conversational practice, while administrators emphasized curriculum-aligned assessment. We offer design recommendations for voice-enabled chatbots in low-resource multilingual contexts, highlighting the need for more intelligible speech output for non-native learners, one-tap interactions with simplified interfaces, and actionable analytics for educators.  Beyond language learning, our findings inform the co-design of future AI-based educational technologies that are socially sustainable within the complex ecosystem of low-resource schools.
\end{abstract}





\maketitle

\section{Introduction}English remains a gateway to academic progression and labor-market opportunity in India, yet many government and low-fee private schools struggle to provide sustained opportunities for spoken practice. Large class sizes, limited exposure to proficient oral models, and teachers with similar proficiency levels as their students constrain interactional learning \cite{ChattopadhayTamo2017,Winson2023}. Infrastructural deficits compound these pedagogical challenges: only 57.2\% of Indian schools have functional computers and 53.9\% have internet access, making it difficult to create sustained, technology-enabled practice environments ~\cite{MinistryofEducation2024}. 
These constraints matter because the economic returns to spoken English proficiency are sizable, especially for youth entering customer-facing and service work \cite{AzamMehtabul2013}. With generative AI increasingly reducing the cost of polished written English \cite{Noy2023,Fang2023}, the relative value of spoken English is only rising. In short, India’s multilingual English-medium classrooms face a structural undersupply of opportunities to speak English, precisely the kind of practice that supports learning and confidence-building.    
Our study sits at the intersection of HCI for Development and second-language pedagogy, guided by two premises: First, in low- and middle-income school systems, generic technology deployments often yield weak or null effects, whereas level-appropriate, teacher-aligned interventions can produce sizeable gains \cite{Muralidharan2019,Rodriguez-Segura2022,Hennessy2022,UNESCO_edu_report}. Second, conversational agents show promise for oral proficiency by supporting turn-taking and contingent feedback, though challenges remain in reliability and feedback quality \cite{Du2024,Wu2024,Zhang2023,Wang2024}. These insights align with established accounts that emphasize negotiation of meaning, ``pushed output'', and low-risk practice environments for second-language acquisition \cite{Long1996,SwainM1995,Nation2003,Macaro2018,Macintyre1998}. 

Together, these dynamics foreground design and deployment challenges for affordable schools in India: sustaining robust voice interactions under intermittent connectivity, coordinating classroom–home technology use within rigid timetables, and offering teacher-facing analytics that aid rather than encumber practice.

Against this backdrop, we conducted a six-day, multi-stakeholder field study of a prototype spoken-English chatbot in four low-fee schools in Delhi. Students interacted with the chatbot daily, while we gathered rich qualitative data through observation and interviews with students, teachers, and principals. 

We adopted an interpretivist, multiple-case qualitative design \cite{Kamal2006}, structured into two phases: (1) real-time observation with immediate feedback (Day 1), where students were introduced to the chatbot and engaged in a supervised session followed by a feedback discussion; and (2) extended use with delayed feedback (Days 2–6), where students continued to use the chatbot daily while researchers observed interactions and collected reflective feedback. This approach allowed us to examine not only immediate usability and reception, but also how perceptions evolved through repeated use.

We focus on three research questions:
\begin{itemize}[nosep]
    \item[]RQ1: How do students, teachers, and principals, in a low-resource, multilingual school environment, experience a voice-based English-learning chatbot?
    \item[]RQ2: What technical and pedagogical factors enable or hinder student engagement with the chatbot?
    \item[]RQ3: What design adaptations could optimize the chatbot’s usability and educational value in such contexts?
\end{itemize}

In the process of answering those three questions, we make the following contributions:
\begin{itemize}[nosep]
    \item An empirical multi-day, multi-stakeholder purposive maximum-variation sampling field study of a voice-first English practice chatbot across four low-resource Delhi schools, detailing confidence trajectories and turn-taking frictions unique to multilingual classrooms (N=23 students, 6 teachers, 5 principals).
    \item Specific design knowledge and tensions that enable sustainable adoption and support in environments where reliability cannot be assumed. 
    \item A checklist for field implementations of educational chatbots in resource-constrained multilingual communities and pathways to integrate with national platforms. 
\end{itemize}

\section{Related Work}

\subsection{Conversational Chatbots}
Conversational agents have become one of educational technology’s most rapidly adopted strands. A recent systematic review found that the research on educational chatbots has grown steadily, with language learning emerging as the single largest application domain \cite{Hwang2023,Fu2024,Labadze2023,Laun2025}. Parallel meta-analyses similarly report robust but heterogeneous learning benefits \cite{Okonkwo2021,Wang2024}. More recent empirical work reinforces this trend: for example, AI chatbots have been implemented as conversation partners in EFL classrooms to increase speaking opportunities and confidence \cite{Yang2022}, and studies of self-directed chatbot use show that learners leverage conversational agents to practice communication strategies and regulate their own learning processes \cite{KangSung2024}.

From a theoretical standpoint, chatbots can be framed within established learning theories. In sociocultural terms, they function as tools in the learner’s Zone of Proximal Development (ZPD) \cite{VYGOTSKY1978} by offering contingent scaffolds (i.e., support tailored to the learner’s current level). Activity Theory provides another lens, focusing on how the chatbot mediates language learning activity as part of a broader system \cite{Cheng2024}, where the chatbot serves as a mediator in the learning process. Chatbots operationalize ZPD by providing contingent scaffolding \cite{Wood1976}, while Activity Theory frames human-chatbot-task mediations \cite{Engestrm2016}. Recent design-oriented work further illustrates how pedagogical structure can be embedded into chatbot systems; for instance, curriculum-driven approaches synthesize conversational data to align chatbot interactions with instructional goals \cite{Li2024}, reflecting a growing emphasis on integrating chatbots more deliberately into educational ecosystems.

Within educational applications, language learning remains the dominant testbed for chatbots. Conceptual reviews \cite{Huang2022,Jeon2023} and three independent meta-analyses \cite{Zhang2023,Wang2024,Wu2024} converge on a moderate mean effect (g $\approx$ 0.50) for chatbot-assisted language learning. These effects are moderated by factors such as the type of language task and interface modality used. Voice-based agents merit special attention. A recent meta-synthesis of 57 voice-chatbot studies \cite{Li2025} found that using voice can make interactions feel more authentic for learners. Still, it also introduces higher speech-recognition difficulties, especially for young and non-native speakers.

\subsection{Chatbots in the Global South}
Clearly, educational technology (EdTech) initiatives have not uniformly solved the problem of teaching spoken English and providing avenues to practice it. Device-only or content-push programs show mixed or null effects without strong pedagogy, while targeted, level-appropriate tools can produce large gains \cite{Cristia2017,Rodriguez-Segura2022,Cueto2024,Muralidharan2019}. Recent policy syntheses caution that technology should be instructionally purposeful and deployed on learners’ terms rather than as a one-size-fits-all solution \cite{UNESCO_digital_platforms}. Voice interfaces are salient for spoken L2 (second-language) development because they afford real-time turn-taking and feedback. Focused reviews of chatbots for speaking report gains in interaction, motivation, and oral performance, while also noting issues of reliability and feedback quality \cite{Du2024,Wu2024}. At the same time, the speech pipeline introduces distinctive frictions: automatic speech recognition (ASR) struggles disproportionately with children’s voices, non-native accents, and code-switched tokens; text-to-speech (TTS) choices around speech rate and accent shape learners’ comprehension and comfort \cite{Jeon2023,Sullivan2022,Chi2022,Ngo2024}. These constraints are precisely those encountered in Indian English medium classrooms where Hinglish and local accent varieties are common.

Notably, one recent randomized evaluation in Nigeria found LLM-based tutoring improved English outcomes for secondary students \cite{DeSimone2025}, while a focused review on speaking practice reports positive effects on interaction, motivation, and oral skills \cite{Du2024}.  In India, scaling efforts such as ConveGenius SwiftChat---reportedly reaching tens of millions of learners and multiple languages---illustrate demand and feasibility for chat-based learning at a population scale, even as peer-reviewed evaluations of school-based, voice use remain limited \cite{Amazon2023}. The empirical base remains skewed toward high-resource settings, with few studies attending to bandwidth variability, accent/intelligibility, or multilingual code-switching typical of Indian classrooms \cite{Hennessy2022,Rodriguez-Segura2022,Jeon2023}. 

\subsection{English as a Second Language}
A language‑learning perspective clarifies when and why voice‑based systems are likely to help. Interaction‑rich tasks create opportunities for negotiation of meaning---learners can request clarification, hear reformulations, and adjust their utterances in response to contingent feedback \cite{Long1996}---while opportunities for pushed output prompt learners to notice gaps and restructure interlanguage \cite{SwainM1995}. Participation depends not only on competence but also on willingness to communicate (WTC); designs that minimize social risk and provide immediate uptake cues can raise situational WTC and make learners more likely to attempt to speak \cite{Macintyre1998}. At the same time, systems should respect cognitive load limits by keeping lexical/syntactic demands within level‑appropriate bands and by offering brief, toggleable L1 (first-language) scaffolds exactly at points of breakdown, so comprehension is supported without displacing L2 interaction \cite{Sweller1988,Hall2012,Garca2014,Macaro2018,Macaro2020,Polio1994,Turnbull2001}.

In English‑medium classrooms, teachers' variable spoken proficiency and large classes narrow chances for extended learner turns and spontaneous negotiation of meaning, often prompting reliance on text‑first routines \cite{ChattopadhayTamo2017,Winson2023,Hennessy2022}. Learners and teachers routinely deploy code‑switching/translanguaging to maintain joint attention and ensure task comprehension \cite{Hall2012,Garca2014,Macaro2018}. Empirical syntheses suggest benefits when L1 use is selective and purpose‑bound---for task set‑up or clarification---without displacing L2 interaction \cite{Nation2003,Macaro2020,Polio1994,Turnbull2001}. These patterns imply concrete design choices for voice agents: toggleable L1 supports that appear and recede, and branching prompts that can clarify, recast, or repeat without teacher mediation.

The practical target in classrooms is comprehensibility rather than a native‑like accent; settings that favor moderate speech rate, stable prosody, and locally familiar varieties reduce processing burden and improve listener understanding \cite{Munro1998,Derwing2005,Setter2017}. Together, these principles specify a voice design that is contingent, low‑risk, cognitively manageable, and tuned to the speech features that matter most for successful classroom communication.

Critically, speech technology choices shape whether these pedagogical affordances are realized. Learners attend more to comprehensibility than nativeness; slower speech and locally familiar accents improve intelligibility and reduce processing load \cite{Munro1998,Derwing2005,Setter2017}. Automatic speech recognition (ASR) can enable feedback on pronunciation and fluency, but performance degrades for child voices, non-native accents, and code-switched tokens common in Hinglish, introducing turn-taking friction \cite{Ngo2024,Jeon2023,Sullivan2022,Chi2022}. For schools with intermittent connectivity and shared devices, single-tap microphone interaction, accent-matched text-to-speech (TTS), and robust ASR for proper nouns and loanwords are not cosmetic choices---they are prerequisites for equitable access to voice practice.   

\subsection{EdTech in Low-Resource Educational Contexts}
Work on educational technology in low-resource contexts reflects longstanding HCI4D (Human–Computer Interaction for Development) concerns around infrastructure, access, and the socio-technical conditions that shape technology use. In rural India, Akkara et al. \cite{Akkara2020}  describe a mobile-assisted language learning (MALL) intervention with first-generation English learners in government schools. Even brief, structured phone-based activities increased learners’ exposure to English, yet remained tightly constrained by connectivity and device access. Raj and Tomy's study with 149 rural Tamil Nadu college students\cite{Raj2024} similarly shows that mobile apps can support listening comprehension gains, but only when learners have reliable device access and content designed for low-resource settings.

Recent work explores how AI-based tools interact with the realities of Global South classrooms. Chatbot-assisted interventions in Saudi secondary schools have shown that AI-driven conversational systems can enhance spoken English and student confidence in what the authors characterize as low-resource EFL environments \cite{Alenezi2025}. Extending this perspective, Verma et al. argue that Child–Computer Interaction with AI systems in the Global South must explicitly account for infrastructural fragility, sociocultural expectations, and inequities in access \cite{verma2025exploring}. These studies underscore the need to design for both pedagogical value and the material conditions under which learners and teachers operate.

At the system level, large-scale EdTech deployments in low- and middle-income countries  reveal persistent cost and sustainability constraints in public school systems. In Honduras, Bando et al. \cite{Bando2017} showed that replacing printed textbooks with laptops in high-poverty elementary schools produced no significant difference in learning outcomes, although laptops could become cost-effective if they substituted enough textbooks, underscoring that hardware alone is not a pedagogical silver bullet. Paterson \cite{paterson2007costs} similarly documents that the total cost of ownership for school ICT in Botswana, Namibia, and Seychelles remains high, with recurrent costs and maintenance often underestimated by policymakers, echoing HCI4D critiques of techno-solutionism and hidden support burdens.

Teacher-facing interventions in low-income Indian schools further highlight these tensions. Smartphone-based programs can extend instructional resources and peer support but also introduce new burdens and techno-stress for teachers with limited time and technical capacity \cite{Varanasi2020,Varanasi2021}. In contrast, Naik et al. \cite{Naik2020} report positive learning gains from a technology-aided teaching program across 1,823 rural government schools in India, where a low-cost digital intervention replaced one-third of regular classroom instruction without intensive teacher retraining. Together, these findings illustrate a consistent HCI4D principle: technology can support learning in resource-constrained systems only when aligned with existing practices and designed with attention to cost, infrastructure, and teacher capacity. These same considerations shape the feasible adoption of a voice-based chatbot in our partner schools.

\section{Method}
\subsection{About the Chatbot}

\begin{figure*}[t]
  \centering
  \includegraphics[width=\textwidth]{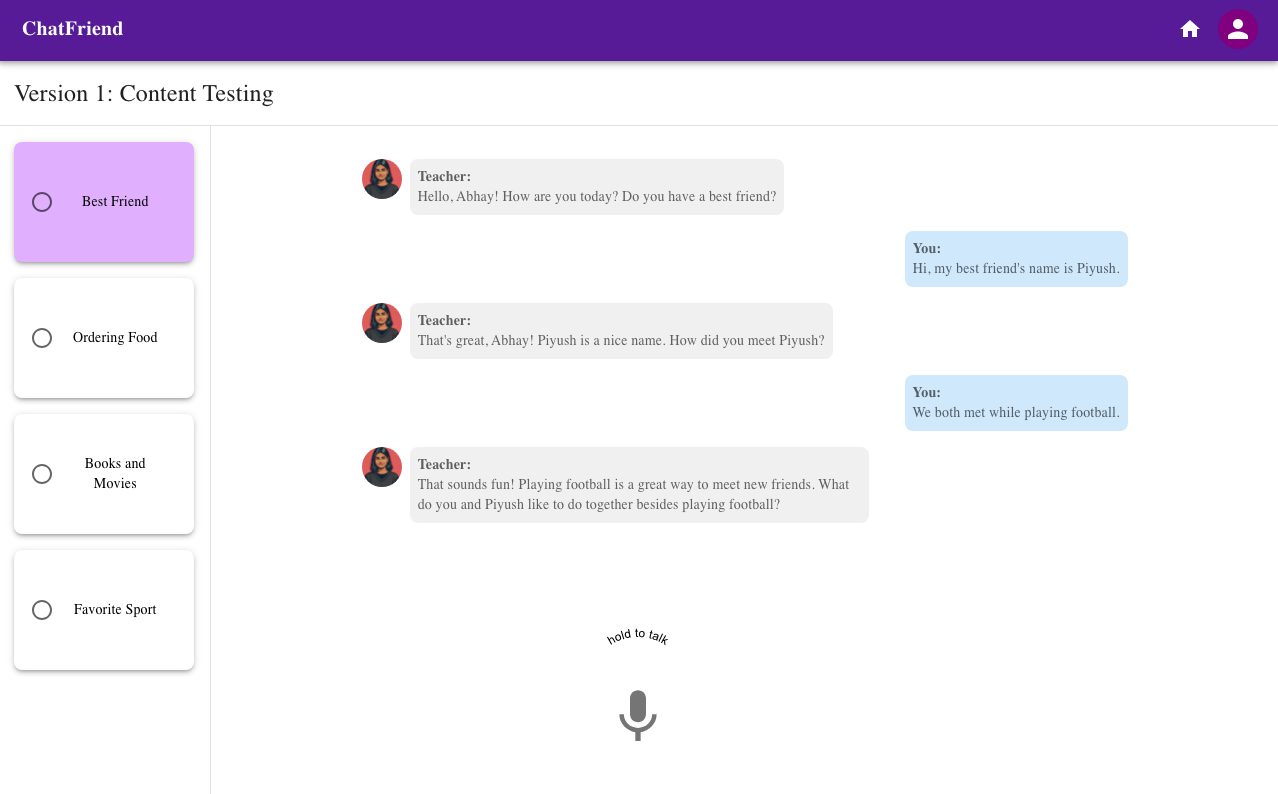}
  \caption{ChatFriend Interface--- Student participants interacted with the ChatFriend web application on research team-provided Android tablets. The left panel displays available conversation topics (e.g., ``Best Friend'', ``Favorite Sports''), and students select one to begin a session. ChatFriend initiated the conversation, and students responded by tapping/pressing the on-screen microphone button to record audio. The system replied with synthesized speech, while text transcripts of both student and system turns appeared in the scrolling chat history on screen.}
  \Description{ChatFriend Interface– Student participants interacted with the ChatFriend web application on research team-provided Android tablets. The left panel displays available conversation topics (e.g., "Best Friend", "Favorite Sports"), and students select one to begin a session. ChatFriend initiated the conversation, and students responded by pressing and holding the on-screen microphone button to record audio. The system replied with synthesized speech, while text transcripts of both student and system turns appeared in the scrolling chat history on screen. }
  \label{fig:interface}
\end{figure*}
We developed ChatFriend, a voice-based chatbot prototype designed to help students practice conversational English. The web-based application (Figure \ref{fig:interface}), built with React, is hosted as a static site on AWS S3 and delivered globally via CloudFront for fast performance. Students can engage in spoken dialogues on topics drawn from their daily lives and school materials (e.g., discussing a best friend or a favorite sport). Interaction is voice-first: users press a ``hold-to-talk'' button to speak. Their audio is transcribed in real-time using Whisper-1 \cite{openai_whisper_2022}, and the text is processed by a GPT-4o-mini model \cite{openai-gpt4o-mini-2024}. Full-duplex voice-based interaction was not publicly available at the time of this study. A custom prompt, informed by the conversation topic and student profile, generates a streaming response (see Appendix \ref{prompts}). This reply is first vetted by OpenAI’s Moderation API \cite{openai_moderation_api} to filter unsafe content before being converted into speech via Google Text-to-Speech and streamed back to the user.
To support comprehension, the interface displays a real-time transcript of the entire conversation. Furthermore, to enhance accessibility in multilingual contexts, the system includes a bilingual support feature. Students may ask clarifying questions in their native language (e.g., Hindi), to which the bot responds by providing English translations or explanations.

\subsubsection{User Journey Walkthrough}
Before each session, teachers completed a brief sign-up step on behalf of students, entering the student’s name and grade level. When students received the tablet, they were taken directly to the main interface (Figure \ref{fig:interface}), where the default conversation topic (``Best Friend'') was preselected. ChatFriend initiated the dialogue with a spoken prompt accompanied by an on-screen text bubble. To respond, students pressed and held the microphone button while speaking and released it when finished. If a student replied in English, the real-time transcription appeared on the right side of the chat history; if they instead spoke in Hindi, the system displayed an English translation in the same location. The chatbot's reply was always delivered in English through synthesized speech and text, ensuring consistent exposure to the target language while still supporting clarification in the student’s native language. Students could switch to a new conversation topic at any time, which would end the current exchange and launch a fresh dialogue sequence.

\subsection{Study Setting}
We conducted the field study in four affordable private schools located in the Badarpur and Shakti Vihar neighborhoods of Delhi, India. Contrary to the ``elite'' stereotype, the private schools in our study are low-fee, neighborhood, often English-medium institutions serving low- to middle-income families \cite{Kingdon2020}. In Andhra Pradesh, a two-stage randomized controlled trial (RCT) showed these schools paid teachers under one-sixth of public salaries and spent less than one-third per pupil while producing comparable Telugu/math outcomes \cite{Muralidharan2015}. This setting is analytically attractive because baseline English proficiency is low (e.g., ASER 2022 reports ~25\% of Grade 5 students can read a simple English sentence \cite{aser_2022}), and parents strongly demand English-medium provision in affordable private schools. We also included a fifth school for added perspective (only the principal of that school was interviewed to broaden the administrative insights). We used purposive maximum-variation sampling to select schools that differed in management type and digital infrastructure to ensure diverse contexts. 

The primary language of most students and teachers was Hindi. Although English was designated as the official medium of instruction, in practice this largely translated into the use of English-language textbooks, while classroom teaching was conducted predominantly in Hindi. As a result, students had limited opportunities to engage in English conversation within the school environment, and the English language proficiency of the student body was fairly low.

\subsection{Participant Recruitment}
We first secured institutional buy-in for the field study by presenting ChatFriend to school administrators as a pedagogical tool specifically designed to develop conversational English skills. We framed it as a complement to existing curricula, which often place considerable emphasis on grammar, reading comprehension, and written expression but offer few opportunities for authentic speaking practice. School principals were briefed in person and via follow-up communication, and each formally agreed to participate. 

English teachers then distributed opt-in consent forms to all Grade 7 and 8 students. 
Across the four participating schools, each of which had multiple Grade 7 and 8 classrooms (approximately 20–45 students per class), we first randomly selected two classrooms per school from all eligible Grade 7 and 8 classrooms. From the rosters of these selected classrooms, all students who returned signed parental consent were entered into a simple random sampling procedure using a computer-generated random number list. Our goal was to recruit approximately 20 students in total, with the target roughly distributed across the two grades; because Grade 8 students demonstrated stronger ability to communicate with the bot during the pilot, we slightly oversampled Grade 8 students when feasible. No students were excluded on the basis of English proficiency or academic performance, and teachers did not nominate or screen students beyond distributing and collecting consent forms. This process yielded 23 randomly selected students.
Participating teachers and principals were recruited once their schools joined. No consenting participants dropped out. 

\begin{table*}[t]
\centering
\begin{tabular}{p{2.5cm} p{2.5cm} p{6.5cm} p{3.5cm}}
\toprule
\textbf{Stakeholder} & \textbf{Sample size} & \textbf{Inclusion criteria} & \textbf{Role in study} \\
\midrule
Students & 23 (13 female, 10 male)& 17 were in Grade 8, and 6 were in Grade 7; Regularly enrolled, parental consent, no prior chatbot exposure& Primary users of the bot\\
English teachers& 6 (5 female, 1 male)&Taught Grades 6-8 in participating schools&Interviewees, observers of student-bot interactions\\
Principals&5 (3 female, 2 male)&Head of participating schools&Key informants on feasibility and policy fit\\
\bottomrule
\end{tabular}
\caption{Description of the stakeholder groups engaged in the field study}
\Description{Table describing stakeholder groups in the field study. 
Three groups participated: Students, English teachers, and Principals. 
There were 23 students (13 female, 10 male), mostly in Grade 8 with some in Grade 7, included if they were regularly enrolled, had parental consent, and no prior chatbot exposure. 
They served as primary users of the chatbot. 
There were 6 English teachers (5 female, 1 male), all teaching Grades 6–8, who acted as interviewees and observers of student–bot interactions. 
There were 5 principals (3 female, 2 male), heads of the participating schools, who provided insights on feasibility and policy alignment.}
\label{tab:participants}
\end{table*}

To strengthen the credibility of our findings, we engaged all three stakeholder groups for data triangulation. Their composition and roles are detailed in Table \ref{tab:participants}.
The study was was approved by the Institutional Ethics Committee of Ashoka University (Approval No. 24-F-10019-Sharma). Consent was obtained at multiple levels: institutional consent from schools, written consent from teachers and principals, and written assent from students, alongside parental consent facilitated by the schools. Schools were provided with written information about the study to share with parents. Participation was voluntary, with students free to skip questions or withdraw at any time; chatbot use and feedback had no impact on grades. No individual compensation was provided, though each school received books for its library. All data were anonymized, encrypted, and accessible only to the core team. Participants (and parents of minors) could decline audio recording, though all agreed. 

\subsection{Data Collection}
\subsubsection{Procedures}
Before data collection, all field facilitators received comprehensive training in two areas: (1) systematically observing and recording student-chatbot interactions using a provided checklist, and (2) conducting productive interviews and probing for feedback with students, teachers, and principals. (See the checklist and discussion guides in Appendices ~\ref{student_interview}--\ref{observer_checklist}). Throughout the deployment, the research team held regular debriefing sessions to cross-check interpretations and emerging findings.

\textit{\textbf{Data Collection Phase 1 (Day 1)}---Initial Student-Bot Interactions with Immediate Feedback:}
On the first day at each school, we conducted two back-to-back sessions with the students: 
\begin{itemize}[leftmargin=*]
    \item \textbf{Initial chatbot session (15–20 minutes, supervised): } Facilitators began by briefly demonstrating the chatbot's features. To start a session, a facilitator would log into a student's account, select the first task, and then hand the tablet to the student. After the student completed the conversation, the facilitator would select the next task and return the tablet. Each student interacted with the chatbot for 15–20 minutes under close observation. Facilitator’s observations focused on three areas: interface navigation, engagement levels, and any difficulties encountered. Facilitators documented their observations in real-time using handwritten or checklist-based notes.

    \item \textbf{Immediate feedback discussion (5–10 minutes):} Following the students' initial interaction with the chatbot, we facilitated a short debrief. Using open-ended questions, we gathered their first impressions, including what they liked, disliked, and any problems they encountered. This guided discussion allowed students to reflect on the experience while it was fresh, and for us to capture their initial reactions in their own words.
\end{itemize}

\textbf{Interviews with School Principal and Teachers:}
On Day 1, we interviewed the school principal for 35–45 minutes. This discussion covered higher-level issues, including the chatbot’s alignment with school priorities, infrastructure requirements, and its perceived value for student learning.

We then interviewed the English teachers after they observed the first student-bot interaction session. These one-on-one semi-structured interviews, which lasted 30–40 minutes, focused on the teacher's perspective regarding classroom integration, student engagement, and any necessary support or training.

\textit{\textbf{Data Collection Phase 2 (Days 2–6)}---Extended Use with Delayed Feedback:}
Over the next five days, students used the chatbot in brief daily sessions at school. A facilitator was present each day to ensure the sessions ran smoothly, encourage participation, and quietly observe interactions. Each student typically spent 10–15 minutes daily with the chatbot.

The facilitator noted any emerging patterns, such as increasing student confidence or conversation length, and tracked technical issues. Students were encouraged to report problems after each session, all of which were documented. This extended period allowed students to become more familiar with the system and provide more considered feedback beyond the novelty of the first interaction. At the end of each daily session, students were asked an open-ended, unprompted question for feedback, which was used for qualitative analysis.

\subsubsection{Instruments}
Data collection was conducted using researcher-provided Android tablets for student-chatbot interactions. All student discussions and teacher/principal interviews were audio-recorded with consent. These recordings were later transcribed verbatim, with any Hindi or code-switched speech translated into English for analysis. They were analyzed in conjunction with facilitator observation notes to ensure the accurate capture of details and quotes. Personal identifiers were removed from transcripts and observation notes, and participants were assigned pseudonymous IDs (e.g., ``4-4-1S-4'') to track inputs across days and data sources without revealing identities. 

\subsubsection{Teacher \& Principal Roles}
During the deployment, teachers and principals played clearly delimited roles. Principals functioned as institutional gatekeepers, approving participation, helping to schedule sessions, and providing school-level context in interviews; however, they did not observe or intervene in individual student–bot conversations. English teachers coordinated logistics (e.g., escorting students to the study area) and, when present during sessions, acted as non-intervening observers rather than co-teachers. They refrained from correcting students’ language or coaching responses so that all utterances during the chatbot sessions would reflect students’ spontaneous use of English. Their primary research role was as interview participants and as interpretive informants on how the chatbot might fit into their existing pedagogy.

\subsubsection{Facilitator Roles}
Facilitators were members of the research team rather than school staff. During student–chatbot sessions, their role was deliberately constrained: they were instructed not to correct students’ English, suggest answers, or otherwise coach content. Instead, they focused on setting up devices and accounts, quietly monitoring interactions, and intervening only to resolve technical issues (e.g., microphone problems, log-in difficulties) or to offer brief, non-instructional encouragement (such as reminding a student that they could try again if the bot misheard them).
During these sessions, facilitators stayed in the room but refrained from providing linguistic help or evaluating performance; their support was limited to troubleshooting the interface and offering neutral prompts to continue (e.g., ``You can speak again'' if a student fell silent). Facilitators’ observations focused on three areas—interface navigation, engagement levels, and any difficulties encountered—which they documented in real time using handwritten checklist-based notes.

\subsection{Analytic Approach}
We followed a multi-step qualitative analysis process involving three members of the research team.
Our analytic orientation was primarily interpretivist and inductive: rather than applying predefined categories, we allowed codes and themes to emerge from the data. Because the work involved multiple analysts, we used intercoder agreement as a pragmatic check on the clarity and consistency of our collaboratively constructed codebook, not as a claim that a single ``correct'' interpretation exists.
\begin{enumerate}[leftmargin=*]
    \item Familiarization – The research team read all transcripts and field notes twice and wrote margin memos.

    \item Open coding – Two analysts independently conducted inductive, line-by-line coding on an initial subset of approximately half of the data, generating provisional codes capturing recurrent ideas, emotions, and interactional patterns (e.g., ASR mishearing, difficulty navigating the interface, confidence surge, assessment appetite). The analysts then compared and merged code lists, drafted explicit inclusion–exclusion criteria, and added example excerpts for each code. Using this shared codebook, they independently recoded the same subset and calculated raw percentage agreement on the presence of each code within segments; agreement exceeded 90\% after reconciliation. Remaining discrepancies were discussed to clarify boundaries between adjacent codes and to determine whether new codes were warranted. The finalized codebook was then applied to the full dataset, with a third researcher available to adjudicate any remaining ambiguous cases.

    \item Axial \& temporal coding – Codes were clustered into categories (e.g., microphone interaction user experience,
    ASR mishearing, confidence shift). For students with $\geq$3 sessions, matrices plotted codes against session numbers to visualize the trajectory.

    \item Theme construction \& triangulation – Themes were refined through constant comparison across data sources and stakeholder groups. We explicitly searched for negative cases—segments or participants that contradicted or complicated emerging patterns (e.g., students whose confidence decreased over time). Rather than excluding such cases, we used them to refine theme boundaries (e.g., distinguishing fragile confidence trajectories from uniformly positive ones) and to nuance interpretations in the Results and Discussion.

\end{enumerate}

\section{Results}
\subsection{Pre-engagement Reception of ChatFriend}
When introduced to ChatFriend, a number of schools expressed enthusiasm, noting that conversational fluency is a critical dimension of language proficiency that is often underdeveloped in classroom settings. In these cases, ChatFriend was welcomed as an innovative supplement to formal instruction that could increase students’ confidence and ease in spoken English. School administrators further appreciated the chatbot’s potential as a pedagogical aid and mentioned interest in the bot discussing science and other lessons with students. However, other schools were more cautious or resistant. Several cited structural barriers, such as insufficient internet infrastructure, limited classroom time within an already overextended curriculum, and the additional demands placed on teachers. 

Teachers also raised concerns about attention and appropriate phone use, particularly when considering where the chatbot should fit into students’ daily routines. For instance, while one teacher felt that home use might be more feasible because ``using the bot in the classroom wouldn’t be possible due to the packed syllabus and issues like absenteeism'' (Teacher T1), others worried that assigning the chatbot as homework could inadvertently encourage over-reliance on phones: ``Students might become too dependent on phones. I’d prefer they use it in the classroom'' (Teacher T5).

School leaders similarly mentioned concerns over students’ engagement with online technologies. In particular, the prospect of learners interacting with the internet without adequate supervision was perceived as problematic, with stakeholders expressing apprehension about issues of safety, distraction, and appropriate use. 
These concerns extended to both technical and social conditions for safe engagement. One teacher highlighted that students ``need personal and private space for using it without hesitation'' (Teacher T4), underscoring that privacy and technical support shape whether students can participate comfortably. Others anticipated that parents might support home use but would not always be positioned to provide effective oversight: ``Parents wouldn’t have any issue … but guidance from parents might be a challenge'' (Teacher T3).

At the same time, teachers recognized that ChatFriend could offer certain forms of social safety that classroom interactions do not. As one teacher observed, ``The bot offers a safe and non-judgmental environment for students who may hesitate or lack the confidence to practice their conversational skills'' (Teacher T2). Another noted that this sense of safety might even enable students to raise topics ``we might hesitate to talk about with others'' (Teacher T1). Together, these perspectives show that teachers and principals saw both risks in unsupervised technology use and opportunities to create a psychologically safe space for language practice.

Together, these perspectives address RQ1 by showing that early experiences of ChatFriend were shaped not only by its perceived pedagogical value, but also by practical concerns around supervision, safety, and appropriate technology use, which in turn influence whether and how engagement is viewed as feasible (RQ2).

\subsection{Student Background and Motivations}
The student sample comprised 23 students across four schools in Grades 7 and 8, whose primary language is Hindi. Opportunities for spoken English were confined mainly to school settings at best, while provision varies across schools---at least one student reported not having English classes, which helps explain why learners seek extra practice opportunities. Outside of school, English exposure or practice is extremely limited for these students. 
Students overwhelmingly perceive English as an important skill set and have strong instrumental motivation to learn. In Day 1 interviews, virtually all respondents ($\approx$95\%) framed English as essential for communication, job interviews, or future mobility, with about 60\% also citing international travel. Students who aspired to medical (35\%) and teaching (15\%), or careers stressed English as a prerequisite for success.
Students mentioned opportunity for English speaking practice at school is limited, and engagement hinges on the teaching style and interruptions like exams/parent-teacher meetings. Only a minority of the group (22\%) described themselves as feeling comfortable with or enjoying speaking in English, while most students in the group (78\%) conveyed anxiety and nervousness with spoken English, citing fear of mistakes and limited vocabulary.
\begin{quote}
    ``I hesitate when speaking English because I fear getting it wrong.'' —Grade-7 student (ID S-7-12)
\end{quote}

Students' linguistic backgrounds, motivations, and speaking anxieties provide important context for understanding their experiences with ChatFriend (RQ1) and help explain factors that shape initial and sustained engagement with a voice-based chatbot (RQ2).

\subsection{First-encounter Experience}
Students' first encounters with ChatFriend were marked by initial hesitation that, for many, shifted toward greater comfort over time, offering insight into how early experiences with conversational agents evolve in low-resource learning contexts (RQ1).

On Day 1 of the study, facilitators noted cautious first use as students had their first interaction session with the chatbot. Students often produce short, hesitant replies to chatbot messages. On Day 1, 36\% of the student utterances consisted of no more than 3 tokens (these are often direct responses, such as ``Yes'', ``<friend’s name>'', ``Cricket'', to the bot's questions without further elaboration or engagement). The facilitators also made note of the chatbot's speech misrecognition issues (especially with proper nouns such as friends' names, games, sports figures) appearing in many of the student-bot conversations and creating.
After completing their first interactions with ChatFriend on Day 1, students reflected on their first-encounter experience with the chatbot, which revealed a more nuanced emotional landscape.
Several students described initial discomfort or feeling ``weird'' talking to a non-human interlocutor, especially during early sessions:
\begin{quote}
    ``At first, it felt strange talking to a robot instead of a person.'' —Grade 8 student (ID S-8-06)
\end{quote}
Others reported increased comfort over time, appreciating the bot’s patience and non-judgmental nature, which helped reduce speaking anxiety:
\begin{quote}
``I was nervous, but the bot did not make me feel bad for mistakes.'' —Grade 7 student (ID S-7-14)
\end{quote}

During the day 1 feedback session, students were also asked to rate the task, conversation flow, and interface on a scale of 1 to 5 stars (1 being the least and 5 being the maximum). Despite the initial hesitation of interaction and speech recognition hiccups, students reported predominantly positive responses (task: 43\% 5 stars, 39\% 4 stars, 17\% 3 stars; conversation flow: 48\% 5 stars, 30\% 4 stars, 22\% 3 stars; interface: 43\% 5 stars, 39\% 4 stars, 17\% 3 stars)

English teachers were interviewed, interacted with the bot, and observed student-bot interactions. All six teachers endorsed the bot’s capacity to provide ``non-judgmental'' practice space for shy students. The teachers gave an average rating of 4 for the chatbot tasks. 4.3 for the conversational flow and 4.4 for the format. One teacher specifically remarked that the bot’s ability to register Hindi as English text was a helpful feature for learning correct English expressions. 

These patterns highlight how both technical factors (e.g., speech recognition accuracy) and pedagogical qualities (e.g., non-judgmental responses) shape student engagement (RQ2) and point toward design considerations discussed in relation to RQ3.

\subsection{Extended Use Experience}
\label{student_days}
Observations of students' extended interaction with the chatbot over the next five days revealed a clear pattern of growing confidence. For the 17 students who participated at least five days of the study, the proportion of conversations involving at least one student-initiated question increased from about 36\% on Day 1 to approximately 48\% by Day 4, and 65\% by Day 5. By Day 5, facilitators' notes indicated that 76.5\% of learners predominantly spoke English, 11.8\% primarily used Hindi, and 11.8\% code-switched fluidly. On Day 1 for this same group, the distribution was 29.4\% predominantly English, 64.7\% primarily Hindi, and 5.9\% code-switching. These patterns suggest an evolving linguistic negotiation as learners gain comfort.

Students' self-reports complement observation notes that document longer, more fluent utterances over repeated sessions. Several learners noted improvements in their English vocabulary and speaking confidence attributed to the repeated chatbot interaction. Examples include remembering new words or phrases encountered during dialogue and increased ease in forming sentences:
\begin{quote}
    ``I learned new words like ‘interesting’ and ‘favorite’ by using the bot.'' —Grade 8 student (ID S-8-02)

``Talking with the bot made me more confident to speak in class.'' —Grade 7 student (ID S-7-10)
\end{quote}

\subsubsection{Student Affective Trajectory}
To summarize how students’ affect evolved across sessions, we first generated a session-level affect rating for each student. For each student–session, one researcher read all available narrative fields in the observation checklist, including the Sentiment note, the enjoyment description (``Is the student enjoying the conversation?''), the three items assessing the conversation (brief vs. descriptive responses vs. asking questions), and any optional Feedback, Other, or technical-issue comments. Based on this combined description, they assigned an overall affect rating on a 5-point scale, where 1 indicated highly negative or anxious affect (e.g., fear, frustration, wanting to stop), 2 indicated somewhat negative affect (e.g., boredom, hesitancy, visible nervousness), 3 indicated mixed or neutral affect, 4 indicated somewhat positive affect (e.g., interest, enjoyment, reasonable confidence), and 5 indicated very positive and confident affect (e.g., strong enthusiasm, high confidence, eagerness to continue). These session-level ratings produced an affective trajectory for each student across up to six sessions, which can be visualized as a student-by-day heatmap (see Appendix \ref{heatmap}) and used to characterize typical ``Affect Arc'' patterns.

Using these trajectories, we classified each student into one of four ``Affect Arc'' types. Three students 
showed a \textbf{Confidence surge}, moving from low or mixed affect in the first session to strongly positive, confident engagement that was largely sustained. 
Four students 
exhibited a \textbf{Steady gain}, with affect gradually rising from moderate to consistently high by the final observed session. 
Three others held a steady pattern 
without a gain. 
Four students 
followed an \textbf{Upward then frustration} arc, starting in the neutral-to-positive range, improving mid-week, and then ending with low scores as frustration or fatigue set in. 
Six students 
showed an \textbf{Excited rise then disrupted} arc, where an initial increase into positive affect was later weakened by dips or volatility without necessarily ending at the very bottom of the scale. Two students remained unclassified due to limited participating sessions.
While these arcs capture recurrent patterns, they also revealed considerable individual variability.

\begin{table*}[t]
\centering
\resizebox{\textwidth}{!}{%
\begin{tabular}{p{2.6cm} p{3.2cm} p{4.2cm} p{5.0cm} p{4.2cm}}
\toprule
\textbf{Affect Arc} & \textbf{Day 1} & \textbf{Day 3} & \textbf{Day 5} & \textbf{Learner ID} \\
\midrule
Confidence Surge
&Bored
&Enjoys Hindi-to-English translation
&"Spoke freely"
&1\_4\_1S\_3, 3\_2\_1S\_1,4\_3\_1S\_4\\
\addlinespace[4pt]
Steady gain
&Nervous, very brief answers
&"More comfortable"
&"Felt confident today"
&
3\_3\_1S\_1, 3\_4\_1S\_1, 3\_2\_1S\_2, 4\_2\_1S\_4\\
\addlinespace[4pt]
Upward trend then frustration
&Reads transcript, pace fast
&"Replying confidently"
&Asks meta-questions but is irritated by lag
&
1\_3\_1S\_3,2\_1\_1S\_2,4\_1\_1S\_4, 4\_4\_2S\_4\\
\addlinespace[4pt]
Excited rise then disrupted
&Nervous, did not ask any questions
&"Responding with excitement"
&"I wanted to end the conversation quickly because of misinterpretations"
&
1\_1\_1S\_3,1\_3\_2S\_3,3\_1\_1S\_1, 3\_3\_2S\_1,3\_4\_2S\_1,4\_3\_2S\_4\\
\bottomrule
\end{tabular}
}
\caption{Progression of the Four Types of Student Affective Trajectories towards the Chatbot}
\Description{ 
This table summarizes four types of learner affective trajectories observed across three study days, along with associated learner IDs. The ``Confidence Surge'' group began Day 1 feeling bored, became engaged by Day 3 through activities such as Hindi-to-English translation, and by Day 5 spoke freely. Learner IDs: 1_4_1S_3, 3_2_1S_1, 4_3_1S_4. The ``Steady Gain'' group started nervous and gave very brief answers on Day 1, appeared more comfortable on Day 3, and by Day 5 reported feeling confident. Learner IDs: 3_3_1S_1, 3_4_1S_1, 3_2_1S_2, 4_2_1S_4. The ``Upward Trend Then Frustration'' group initially read transcripts quickly on Day 1, replied confidently on Day 3, but by Day 5 expressed irritation and asked meta-questions due to system lag. Learner IDs: 1_3_1S_3, 2_1_1S_2, 4_1_1S_4, 4_4_2S_4. The ``Excited Rise Then Disrupted'' group was nervous and did not ask questions on Day 1, responded with excitement on Day 3, but on Day 5 wanted to end conversations quickly due to misinterpretations. Learner IDs: 1_1_1S_3, 1_3_2S_3, 3_1_1S_1, 3_3_2S_1, 3_4_2S_1, 4_3_2S_4. Overall, the table conveys how emotional engagement with the chatbot developed differently across learners, illustrating four distinct affect arcs from initial interaction to later sessions.
}
\label{tab:affect_arc}
\end{table*}

Although many learners demonstrated increased confidence and proactive engagement over repeated chatbot interactions, this progression was not uniform. As reflected in the diverse Affect Arcs in Table \ref{tab:affect_arc}, several students experienced setbacks characterized by nervousness, disengagement, or reliance on facilitator assistance. Such fluctuations are often correlated with technical frictions---namely, speech recognition errors, delays in response, and usability challenges with the microphone interface. These findings underscore the fragile nature of ESL students’ confidence development and highlight the need for consistently low-friction, supportive user experiences to sustain learner motivation over time.
\subsubsection{Demographics-based Comparative Analysis}
We examined affective trajectories by gender, but did not find clear systematic differences. Both girls and boys appeared across all four Affect Arc types. Among girls, volatile arcs (``Excited rise then disrupted'' or ``Upward then frustration'') were slightly more common (7 of 13) than relatively stable or improving arcs (``Steady'', ``Steady gain'', or ``Confidence surge''; 6 of 13). Boys showed a more even distribution, with 4 of 10 in volatile arcs, 4 of 10 in stable or improving arcs, and 2 unclassified due to missing session data. Given the small numbers in each group and the interpretive nature of the arc categories, we treat these as descriptive tendencies rather than robust gender effects.

We also explored whether affective trajectories differed by grade (7 vs. 8), but patterns were not strong enough to support firm conclusions. Almost all Grade 7 students (5 of 6) fell into the more volatile ``Excited rise then disrupted'' or ``Upward then frustration'' categories, whereas Grade 8 students were more evenly distributed across volatile and relatively stable/improving arcs (9 of 17 in ``Steady'', ``Steady gain'', or ``Confidence surge''). Given the small numbers in each group and the interpretive nature of our arc categories, we treat these as descriptive tendencies rather than robust grade effects.

\subsection{Lessons on Design Considerations and Challenges}
\subsubsection{Scaffolding in ESL learning}
At the beginning of the field study engagement, both the school principals and English teachers pointed out that students prefer speaking in Hindi even within school and often require the use of Hindi from teachers to support learning, further weakening the already limited English-speaking environment. Teachers echoed this concern, noting that bilingual strategies were often necessary for comprehension but came at the cost of fluency.
Although the school principals were positive that ``with proper teaching, around 50\% (of the student body) could become proficient'', a majority of them estimated only 30–50\% of students can speak in English, and that most students ``lack confidence and hesitate while speaking (English)''.

For this reason, the principals predicted that students would struggle unless the bot spoke more slowly and with an Indian accent. Principals linked chatbot audio intelligibility directly to inclusive learning opportunities, noting that some students may ``switch off the moment they cannot parse the audio.''

Indeed, during the chatbot interactions, students experienced challenges in comprehending the bot messages. Even though ChatFriend was prompted (see Appendix \ref{prompts}) to speak at middle school or elementary levels, post-hoc text analysis on the chatbot’s and students’ utterances shows that, on average chatbot produced English at Flesch-Kincaid \cite{Kincaid1975} 12th grade level. By contrast, students’ English utterances are on average at Flesch-Kincaid 2nd grade level, and consist predominantly of vocabulary (85\%) at the CEFR A1 level (Common European Framework of Reference for Languages), which indicates the most beginner level of English. This gap in English level likely posed barriers to students’ comprehension of chatbot messages and resulted in students' relatively low engagement in conversations. Compared to an average of 29 tokens produced by the ChatFriend per turn, students, on average, only produced 5 tokens per turn.

When prompted to reflect on their experience after extended use of the chatbot, students expressed interest in features such as adjustable audio speed control and choice of bot voice (preferring one with a Hindi accent). Additionally, students wanted an optional Hindi translation toggle to manage ``Hinglish'' episodes, similar to the bilingual support they are used to in their current classroom environment.

Teachers' perspective converged with student experience, as teachers also reported their observations that the long or multi-clause bot messages could overwhelm learners and suggested that ``shorter sentences would help hesitant users.'' Moreover, teachers pointed out that the bot's speech delivery was ``too fast, with an unfamiliar accent.''

\subsubsection{Technical frictions and coping strategies}
Our field deployment revealed several technical friction points experienced by students. These observations yield critical design considerations for voice-based educational tools in similar contexts.
 
First, observations corroborated by both teachers and facilitators indicate that nearly all students experienced Automatic Speech Recognition (ASR) error issues during chatbot interactions. These errors occurred most frequently with proper nouns and Hindi loanwords. The ASR system also struggled with non-standard accents or imperfect pronunciation of English words. For instance, Hindi-accented pronunciations may render school as something like ‘es-chool’ or separate the initial consonant cluster, producing an extra vowel sound (e.g., /\textipa{@sku:l}/
), which often leads to misrecognitions by ASR systems not trained for such phonotactic patterns \cite{Broselow1998}. Furthermore, students often code-switched into Hindi when encountering unfamiliar English vocabulary, intending to leverage the chatbot's bilingual support. However, the system's processing of Hindi and Hinglish inputs was inconsistent; we documented several instances where it either misinterpreted or blocked these phrases. These recurrent ASR failures disrupted conversational flow and contributed to learner frustration, illustrating the unique challenges of deploying such technology in multilingual educational contexts. 

Second, facilitators observed that approximately 35\% of learners required explicit guidance to operate the hold-to-talk microphone button correctly. Teachers confirmed this challenge, as one noted that students ``struggled with the mic button'' (Teacher T4). Many students demonstrated low digital literacy: their daily routines involved limited interaction with digital devices, and only 60\% had prior experience logging into apps such as Instagram or ChatGPT. Notably, 20\% had never logged into any online service and anticipated difficulty navigating a web application. In response, students suggested a simplified one-tap microphone activation and a larger, more prominent icon to replace the hold-to-talk mechanic. This finding highlights the importance of UI design tailored for learners with limited prior exposure to technology.

Finally, facilitators reported network latency or system errors in nearly half of the student-bot sessions, occasionally leading to premature termination. This reveals a critical design challenge for scaling the application beyond its prototype stage: ensuring robust performance under high, concurrent user load.

\subsubsection{Sustaining student engagement}
Despite overall positive trends, facilitators noted a notable subset of students who exhibited persistent hesitation, nervousness, or disengagement during repeated interactions with the chatbot on Days 2-6. Observation notes captured signs of low enthusiasm, including short replies, avoidance of speaking, and requests for facilitators’ assistance. These affective challenges often correlated with technical frictions discussed above, suggesting that while many students grew more comfortable, their experiences were uneven; therefore, tailored support may be warranted for long-term implementation in schools.

During the Day 1 interview, teachers also anticipated dwindling student engagement with extended use. To counteract students’ decreasing attention spans and boredom, teachers suggested incorporating videos, visuals, cartoons, games, jokes, and interactive scoring systems to help maintain their interest.
When asked what additional features and improvements they would like to see in future iterations of the chatbot, students' wish-list aligns with teachers’ suggestions but adds a richer engagement layer. Specifically, students requested coverage of a wider range of topics (food, gaming, careers, science, mathematics) and richer visual effects (emojis or avatars).

\subsubsection{Appetite for formative assessment}
Through our field engagement, we identified teachers and school administrators---alongside students---as key stakeholders in student–chatbot interactions. All principals (5/5) emphasized the value of transcripts or dashboards that could make individual learner progress visible to both teachers and parents. Currently, teachers rely on informal methods to assess spoken English, such as classroom conversations, presentations, and reading activities. However, they noted that students’ fear of making mistakes often discourages participation. Consequently, every teacher (6/6) anticipated instructional benefits from access to chatbot transcripts and learner-level analytics via a concise dashboard. They stressed that such data must be filterable to highlight growth or recurring errors, so as not to add to already heavy preparation demands. While half of the teachers (3/6) expressed openness to using the chatbot as an assessment tool, concerns remained regarding curricular alignment and limited digital infrastructure.

\subsection{Deployment Preferences and Feasibility 
}
\subsubsection{Student Preferences}
 Students expressed a strong preference for home use (65\%), citing the benefits of privacy and reduced peer scrutiny. A smaller fraction preferred in-class use (25\%), while 10\% expressed flexibility. This preference is contextualized by their primary access to technology: most rely on a parent-owned Android phone at home, with few having consistent access to a personal laptop or stable Wi-Fi. As one Grade-8 student (ID S-8-04) shared, ``I'd like to use it at home in my free time on my dad’s phone in the evening.''

\subsubsection{Principal Perspectives}
Principals predominantly advocated for classroom initiation to establish routines and provide supervision. Three schools (60\%) wished to launch the chatbot inside the classroom for orientation, then migrate use to homework once routines are solidified. The prospect of visible learning gains tempered concerns about unsupervised phone use at home. Three principals expected parents to appreciate the chatbot, while two anticipated concerns around unsupervised internet access that may fade with visible learning gains. For at-home practice, parental support was generally seen as limited, with two principals reporting a lack of digital familiarity or a lack of time as constraints to parental engagement.

Infrastructure constraints strongly shaped feasibility. Students’ device and internet access is constrained and shared. Most students only use the internet for homework occasionally; the majority do this on a parent’s mobile phone, with only a few having access to a laptop and stable Wi-Fi. 

Across the five schools, only three (60\%) possessed even a handful of functioning desktop or laptop computers, and just one school permitted students to go online during class time. Principals consequently framed mobile phone compatibility as the sole scalable path for any technology intervention.
\begin{quote}

    ``Without phones, the plan dies---our lab has two working PCs and forty students.'' —Principal P2
    
``Connectivity drops at least twice a period; any tool must survive offline for a bit.'' —Principal P4

\end{quote}

Principals also emphasized staff capacity: Four of five principals (80\%) reported that teachers could dedicate only ``30 minutes, three times a week'' to new activities---any more would exceed staff capacity without additional hires. As a result, principals recommended that the chatbot be used in short, structured sessions rather than as an open-ended daily tool. They argued that integrating brief, scheduled exercises (e.g., during English periods) was more feasible than expecting teachers to supervise longer or more frequent use. Securing teacher buy-in therefore requires that the tool integrate into the curriculum without increasing workload.

\subsubsection{Teacher Perspectives}
Interviews with teachers revealed three distinct perspectives on chatbot deployment. Half advocated for homework to facilitate extended practice, a third preferred dedicated in-class periods to ensure instructional alignment, and the remainder supported hybrid models. This division centered on a key trade-off: concerns about in-class device availability and potential disruptions were weighed against fears of promoting excessive phone use at home. As one teacher (T5) cautioned, ``Not as homework; students might become too dependent on phones.''

Teachers also highlighted the need for curriculum alignment, hands-on training, and feasible integration into existing 35–45 minute English periods. They stressed that any new technology must replace rather than add to current activities. Although half of the schools already run ``smart-class'' video sessions, none currently use digital assessment tools, which the next iteration of the chatbot is expected to introduce. With most (four of the six) teachers having over six years of experience, sustained professional development will be essential.

\subsubsection{Summary}
Taken together, student, teacher, and principal perspectives underscore the need for flexible deployment strategies that account for device access, connectivity constraints, and staff workload. Successful long-term integration will depend on mobile-first design, short structured activities aligned with the curriculum, and adequate teacher support.

\section{Discussion}
\subsection{Design Tensions in Voice-Based Language Learning Chatbots}
By triangulating perspectives across students, teachers, and administrators, our study extends prior work that has largely centered on single-stakeholder, high-resource contexts \cite{Zhang2023,Wang2024,Godwin-Jones2014}. Our field deployment shows that a voice-first chatbot can create a psychologically safe, low-stakes space for spoken English practice, where many students reported growing confidence. Yet these gains proved fragile: the very qualities that made the chatbot valuable---its empathetic responsiveness and openness---were undermined when technical frictions interrupted the flow of interaction. The central design question, then, is not whether voice-based chatbots can support learning, but how they can sustain that support in environments where reliability cannot be assumed. To address this question, we foreground three design tensions that emerged as defining features of this space.

\subsubsection{Empathetic Dialogue vs. Pedagogical Scaffolding
}
Students consistently appreciated the chatbot’s patience and non-judgmental stance, resonating with prior work showing that socially supportive dialogue styles can reduce learners' affective filters and create psychologically safe, low-stakes spaces for practice \cite{Okonkwo2021,Huang2022,Satar2018,Zhai2022,Guo2024}. 

They gravitated toward free-form conversation, appreciating the chance to ask questions and experiment with language outside rigid drills. In contrast, teachers and principals emphasized the need for structured, curriculum-aligned scaffolding and mechanisms to track measurable progress, reflecting sociocultural theories that situate learning within guided practice \cite{VYGOTSKY1978} and broader tensions in educational technology between learner autonomy and institutional accountability \cite{vanLeeuwen2023}. 

Our field data illustrate the difficulty of meeting both needs simultaneously: the bot’s empathetic ``friend-like'' dialogue could encourage speaking, but fine-grained control of its English output was technically challenging, limiting the precision with which practice could be scaffolded. As a result, early struggles occasionally fostered reflection and growth, but more often turned into frustration when unsupported, showing that ``productive failure'' \cite{Kapur2016} requires pedagogical framing to avoid confidence erosion.
Indeed, for sustainable deployment in schools, formal assessments remain necessary to substantiate both researcher-observed and student self-reported learning outcomes. Without mechanisms for reconciling these perspectives---e.g., providing both exploratory dialogue and actionable analytics---the long-term viability of chatbot use in school contexts remains uncertain.
This tension highlights the central design challenge: sustaining the benefits of empathetic, open-ended interaction while ensuring sufficient scaffolding, curricular alignment, assessment support, and technical reliability to convert initial enthusiasm into sustained language practice \cite{Kasneci2023}.

\subsubsection{Confidence Building vs. Fragility under Technical Friction} Learners' self-reports of increased confidence were corroborated by longer utterances and more spontaneous questioning during sessions, consistent with prior meta-analytic findings on the efficacy of chatbot-assisted language learning \cite{Zhang2023,Wang2024}. However, this growth was easily undermined by technical failures. Where network glitches or ASR issues persisted, confidence not only plateaued but sometimes regressed, a phenomenon extending prior accounts of ``novelty-decay'' \cite{Huang2022,Fryer2017}. Prior work has identified ASR accuracy, accent intelligibility, and bandwidth stability as central bottlenecks for voice-based learning \cite{Emara2024,Jeon2023,Tanabian2005}; our findings extend this by showing how, in multilingual and bandwidth-constrained school contexts, even small usability failures can quickly unravel nascent confidence. These dynamics suggest that confidence is less a linear gain than a fragile trajectory---one that requires both empathetic scaffolding and robust technical reliability to sustain.

\subsection{A Checklist for Field Implementations of Educational Chatbots in Resource-Constrained Multilingual Communities}
Our findings highlight multiple design priorities for voice-based educational chatbots in low-resource, multilingual contexts.
\begin{itemize}[leftmargin=*]
    \item  \textbf{Speech and recognition.} Intelligible, accent-tuned text-to-speech with adjustable speed, and ASR capable of handling regional proper nouns and Hinglish code-switching, are foundational for sustaining learner confidence. Even small gains here yield disproportionate improvements in affect and engagement \cite{Munro1998,Emara2024}.
    \item \textbf{Scaffolding for ESL learners.} Sustained engagement requires scaffolding that distinguishes between content level and language level. Topics should align with the interests of middle school learners, while conversations should use simplified English consistent with grade 3–4 proficiency. This dual alignment helps learners focus on meaning-making rather than decoding, lowering frustration and maintaining motivation.
    \item \textbf{Low-friction interaction.} Interfaces should prioritize simple, low-friction interactions that reduce the likelihood of user error, particularly for younger or digitally novice learners. Complementary visual cues and lightweight UI scaffolds can further lower cognitive load.
    \item \textbf{Device and connectivity constraints.} Systems must be explicitly designed for deployment on low-end Android devices and to function under conditions of intermittent or unreliable connectivity. Such provisions align with prior evidence that observations that mobile-assisted learning in developing regions only succeeds when infrastructural and digital literacy challenges are directly addressed \cite{Kukulska-HulmeAgnes2020,Li2013}.
    \item \textbf{Teacher-facing analytics.} Dashboards should prioritize exceptions over transcripts, surfacing actionable patterns such as persistent tense errors or recurrent silence. For instance, highlighting persistent tense errors or recurring silence provides teachers with actionable, low-burden insights. Given teachers’ limited time and capacity, analytics must remain ``minimal but meaningful'' \cite{vanLeeuwen2023} to balance institutional accountability with classroom realities.
    \item \textbf{Equity and ethics.} Finally, design must foreground equity by ensuring accessibility on low-end devices and safeguarding privacy. Localized ASR benchmarks should be adopted to avoid penalizing speakers of underrepresented dialects, and data practices should be made transparent to educators and parents alike. Without such measures, deployments risk reinforcing rather than reducing existing educational inequities  \cite{Williamson2020,UNESCO_digital_platforms,Selwyn2020}.
  
\end{itemize}

\subsection{Integration into School Systems at Scale}
For sustainable adoption, voice-based chatbots must align with the realities of low-fee and government schools. Leveraging existing national digital infrastructure, such as DIKSHA (Digital Infrastructure for Knowledge Sharing, an integrated portal for teacher training by the Indian government), allows integration into familiar platforms for lesson delivery and teacher training.

At the classroom level, short, teacher-supervised onboarding sessions can be embedded within existing English periods, after which practice can extend to privacy-respecting home use. Packaging content as curriculum-aligned ``Speaking Lab'' micro-modules (e.g., 10–15 minutes) would help schools adopt them with minimal disruption to timetables.

To make analytics actionable at scale, data flows should be role-based and exception-driven: teachers receive concise alerts (e.g., persistent tense errors, silence/short-utterance flags), while school leaders can access more detailed reports on demand. This design minimizes planning overhead while enabling targeted support. Because many schools already run ``smart-class'' video sessions but lack digital assessment, these modules can be scheduled in existing smart-class blocks, with lightweight in-app quizzes to seed formative data without requiring new hardware. 

Scalability also depends on teacher professional development (TPD). Micro-TPD modules---delivered through the same digital infrastructure---should demonstrate classroom facilitation, interpreting dashboards, and troubleshooting ASR or scaffolding breakdowns. Clear, parent-facing guides and light-touch community facilitation can further support learners beyond school hours.

By embedding chatbot use within existing digital ecosystems, while minimizing additional workload for teachers and families, integration strategies can increase the likelihood of sustained uptake in resource-constrained school systems. In sum, distributing a curriculum-aligned, mobile-first ``Speaking Lab'' within national platforms, coupled with exception-based analytics and micro-TPD, offers a credible, scalable, and sustainable pathway across government and affordable-private systems.

\subsection{Limitations and Future Work}
Our findings are based on qualitative and observational data, which are subject to interpretive coding and the technological variability of real-world deployment. The reported percentages capture thematic trends rather than precise prevalence.  While triangulation across multiple stakeholder groups strengthens credibility, these limitations underscore the exploratory nature of the research. The study’s short intervention period (six days) limits our ability to observe long-term engagement and sustained learning outcomes, and the absence of pre- and post-interaction assessments prevents definitive claims about learning gains.
A further limitation is that all student–chatbot sessions occurred in a supervised setting with a research facilitator present. Although facilitators' involvement was restricted to technical support and non-instructional encouragement, their presence may still have influenced students' willingness to persist, take risks, or reattempt utterances. Several students also expressed a preference for using the chatbot privately at home, suggesting that engagement and confidence may differ in less supervised or fully unsupervised deployments.

Future work should incorporate standardized speaking assessments with qualitative tracking to better quantify learning outcomes. Longitudinal studies with larger and more diverse samples, combined with objective measures, would strengthen external validity. Controlled experiments manipulating accent and speech rate, as pioneered by \citet{Munro1998}, could help isolate their effects on comprehension. Comparative studies of LLM-based versus hybrid rule-based chatbots would also be valuable to clarify the cost-benefit profile of generative AI in bandwidth-limited settings \cite{Kasneci2023}.
In addition, comparing school-based use with more autonomous home use would help disentangle the effects of the chatbot from those of human scaffolding and social context.

As surfaced in our field study, there are multiple opportunities for technical refinements to optimize chatbot performance within school settings and enhance user experience. These findings highlight the value of future co-design with stakeholders, drawing on the design considerations identified here to guide subsequent iterations.

\section{Conclusion}
This study examined how students, teachers, and principals in four low-fee Delhi schools engaged with a spoken-English chatbot. All stakeholder groups valued the chatbot as a safe, motivating space for practice, and our six-day deployment suggested it could boost students’ confidence, though these gains were fragile in the face of technical barriers. While students preferred open-ended conversational practice, administrators prioritized curriculum alignment and assessment, revealing tensions around long-term adoption. We highlight design challenges and offer insights for creating voice-based educational technologies that balance student engagement, administrative priorities, and infrastructural constraints in resource-constrained, multilingual contexts.
\begin{acks}
 This work was supported by the Penn Global Research Engagement Fund  and The Bill \& Melinda Gates Foundation (BMGF) core grant (INV 026871). The funders had no involvement in the study design, execution, analysis, or reporting of the results.
\end{acks}
\bibliographystyle{ACM-Reference-Format}
\bibliography{chatfriend_ref}

\appendix
\section{Interview Discussion Guide with Student Participant (Day 1)}
\label{student_interview}
\textit{[Introduction]} \\
\textbf{Introductory Questions }\\
1. How old are you? What's something fun you enjoy doing?\\
2. Do you have a favorite subject in school? What makes it interesting to you?\\
3. Is there a subject you find challenging or less enjoyable? What do you think could make it more interesting?\\
4. Do you like learning English? What kinds of things do you learn in your English classes? What did you learn yesterday? \\
5. Do you find it easy or sometimes difficult when you talk in English? How do you feel about speaking English with your friends, teachers, or parents?\\
6. Why do you think learning English is important? How do you think it can help you now and in the future?\\
7. What do you dream of becoming when you grow up? Do you think knowing English will be useful for that? How? \\
8. Do you watch TV shows, movies or videos at home? What kind of content do you enjoy watching? Do you watch anything in English? \\
9. Do you get homework that requires you to use the internet, such as researching information or watching educational videos? How do you use a computer, tablet, parents’ phone, shared phone, etc.? \\
10. Do you have a stable internet connection at home? Is it through Wi-Fi or a mobile device's internet?
11. Do you go to tuition classes after school? What subjects do you study there?\\
\textit{[Introduce Student to the Bot \& Invite them to Interact with it by performing 2-3 tasks]
}\\
\textit{The observer should meticulously document the student interaction, including what the student said, how the chatbot responded, and any instances where the chatbot failed to understand the student or vice versa. Note any moments where the child encountered difficulties.}  \\
\textit{(Note: This note-taking will be an essential component of training field staff collecting data - they will be trained on how to take observational notes, what to look out for, and we will also conduct mock sessions with them) \\
The following checklist can be used to summarize the notes. 
Collect feedback from the student after the interaction. }

\textbf{General Feedback on Interaction }\\
11. What did you think of interacting with the bot? Did you find it different from other interactions you've had, say, in your English classes?\\
12. How did the conversation feel to you? Do you have similar interactions at home or school with teachers, friends, parents or tutors? \\
13. In what ways did the conversation with the bot feel different from your interactions with these people? \\
14. Which task during the interaction did you enjoy the most, and why? Which task did you not like as much and why? \\
15. Did you like the bot's voice? If not, what would you change about it and why?\\
16. Were there particular moments or parts of the conversation that you found interesting or memorable?\\
17. Did you like the format(website)? What did you like/dislike about it?\\
18. Was there anything you disliked about your interaction with the bot? Did you ever feel frustrated, and why?\\
19. If I asked you to rate the following aspects on a scale of 1 to 5 stars (1 being the least and 5 being the maximum), how many stars would you give each? \\
* Tasks \\
* Conversational Style\\
* Format of the bot\\
\textbf{Comprehensibility }\\
20. Did you find the speed at which the bot spoke comfortable, or would you prefer it to be faster or slower? \\
21. Did you find the bot's accent straightforward to understand? Were there any instances where you struggled to understand the accent? \\
22. Did you learn anything new from the interaction? Was there any word or concept you hadn't heard before that you figured out during the conversation?\\
23. The bot is designed to understand Indian accents and the Hindi language, but sometimes it can fail. Do you think the bot successfully understood what you said, or did you find it hard to make it understand you?\\
24. Did you use any Hindi words or phrases while using the bot?  Was the bot able to correctly understand them?\\
25. Could you follow the text that came along with the audio? Did you find it correct most of the time, or only sometimes?\\
26. Did you get an opportunity to ask the bot questions? Did you ask the bot any questions? Was it able to answer the questions? Were you satisfied with the responses? \\
\textbf{User-friendliness} \\
27. Did you find it challenging to use the bot? What specifically did you find challenging?\\
28. Was it easy to use the audio button to speak during the interaction?\\
29. (Have the students log into the chatbot and choose tasks) Did you find it easy to log in and navigate? \\
30. Where would you like to use the bot more— at home or in classrooms? Why?  If at home, how do you think you would use it? Would you use it on your parents' phone or another device?\\
31. Do you have any suggestions for improving the bot's tasks or activities? Anything in particular you would like to talk to the bot about? \\
32. What else would you like to see added/changed to the bot to make it more useful or enjoyable for you?\\
33. Would you share your experience using the bot with your friends and parents today? What would you tell them about it? How would your family react if you shared this bot experience with them? 
\section{Interview Discussion Guide with English Teacher}
\label{teacher_interview}
\textit{[Introduction]}\\
\textbf{Introductory Questions }\\
1. How long have you been teaching? How long have you been teaching in this school? Do you teach any subjects besides English?\\
2. How many classes and students are there in the school? On average, how many students are in each class?  How many English teachers are there? \\
3. Do students have access to technology tools such as computers, tablets, or headphones at school?\\
4. Does the school have internet access?

\textbf{English classes at the school}\\
5. How many English periods do students have, and what is the duration of each period?\\
6. Can you describe a typical English class and what aspects are covered?\\
7. How much time per week is dedicated to oral language skills in your English classes?  How often do students get opportunities to practice speaking in English?\\
8. What activities are conducted in the classroom to improve students’ spoken English? Is there a specific curriculum or program focused on developing English-speaking skills at your school? \\
9. In your opinion, what is the current level of students' English-speaking skills? Do you notice any hesitation or lack of confidence among students when asked to speak in English? If so, what are some common challenges they face? Could you share some examples or anecdotes to illustrate this?\\
10. Are there particular tools or techniques effective for evaluating students' language skills? Could you explain how you assess students' language, speaking and fluency skills? Any method or criteria that you use?\\
11. Do students normally take an interest in English classes? Are there any particular aspects of the class that they seem to enjoy or engage with more? \\
12. Teachers often use various strategies to make their classes more engaging. What activities or methods do you use to keep students interested in English?\\
13. What opportunities do students have to converse in English outside their English classes? Do they practice speaking English in other classes, with friends, at home, or in tuition classes?\\
\textbf{Use of technology}  \\
14. Do you use technology-based tools or audio-visual resources to facilitate language learning in your classes?\\
15. Do you use any apps or digital tools to conduct student assessments? If so, can you tell us more about it and how it helps you? Do you know any teachers who use these tools? \\
16. Do students typically have access to phones/computers/internet at home? Are they given any homework which requires them to access the internet at home?\\
\textit{[Introduce the teacher to the bot \& invite them to interact with it by performing 2-3 tasks] }\\
\textit{Note: Encourage the teacher to test out the bot for students so they may want to try to interact like a student. We also want to test if we can make it part of the curriculum.}\\
\textbf{General Feedback on Interaction \& Engagement }\\
17. What did you think of interacting with the bot? How did the conversation feel to you? Do you think students interact similarly with teachers, friends, and parents at home or school?\\
18. In what ways did the conversation with the bot feel different from the interactions with these people? \\
19. Which task during the interaction did you like the most, and why? Which task did you not like as much and why? \\
20. Were the tasks relevant to the context and appropriate for the grade level? \\
21. Given the decreasing attention spans of students and their tendency to get bored quickly, do you think they will lose interest in these tasks easily? How can we ensure sustained interest among students? \\
22. Did you like the bot's voice? If not, what would you change about it and why?\\
23. Did you like the format(website)? \\
24. Was there anything you disliked about your interaction with the bot? Did you ever feel frustrated, and why?\\
25. If I requested you to rate the following aspects on a scale of 1 to 5 (where 1 signifies the lowest and five the highest), how would you rate each?\\
* Tasks \\
* Conversational Style\\
* Format of the bot\\
\textbf{Comprehensibility} \\
26. Did you find the speed at which the bot spoke comfortable, or would you prefer it to be faster or slower? Do you think most students in your class would find it comfortable?\\
27. Do you think most students in your class will be able to understand the accent? What percentage of students in your class might find it challenging to follow?\\
28. The bot is designed to understand Indian accents and the Hindi language, but sometimes it can fail. Did you try using any Hindi words or phrases? How well do you think the bot understood those Hindi words or sentences? Did it respond appropriately?
29. Did you follow the audio, more text or both? Did you find them correct most of the time, or only sometimes?\\
30. Did you ask the bot any questions? Was it able to answer the questions? Were you satisfied with the responses? \\
31. Do you believe the tasks offer sufficient opportunities/prompts for students to ask questions or give detailed responses? If not, what improvements could be made to enhance these aspects?\\
\textbf{User-friendliness} \\
32. Did you find it challenging to use the bot? What specifically did you find difficult?\\
33. Was it easy to use the audio button to speak during the interaction? \\
34. Do you anticipate that students would find it user-friendly? Do you think they can use it independently without needing adult assistance?\\
35. Could students log in to the chatbot and select tasks themselves?\\
36. How do you see implementing the chatbot? Would you prefer to integrate it into classroom activities, assign it as homework, or both? What challenges do you foresee in implementing it in classrooms versus at home?\\
37. If you were to design an English curriculum with the bot in it, where do you see it fitting in?\\
38. What role do parents typically play in their child's education? Do you encourage parents to participate in their child's schoolwork? Would you support allowing parents to manage their child's bot account to monitor practice and ensure discipline?\\
39. Earlier, we talked about assessment tools. Do you believe the chatbot could assist you in conducting some of these assessments? \\
40. Would reviewing transcripts of student interactions be valuable for evaluating students' performance? How? \\
41. If provided with a dashboard [share a sample dashboard for context] showing students' performance in one place, do you think it would be a valuable tool for assessing students? How do you see it assisting you or simplifying your responsibilities? \\
42. Alternatively, would you prefer the bot primarily as a tool for students to practice and continue to use your current assessment methods?\\
43. If there were a feature for you to add tasks or upload content to the chatbot in digital format, how easy would it be for you to find and upload content? Where do you typically find supplementary content for students besides school textbooks?\\
44. Do you have any suggestions for improving the bot's tasks or activities? Is there anything specific you would like students to discuss with the bot?\\
45. What else would you like to see added/changed to the bot to make it more helpful or enjoyable for the students?\\
\textbf{Perceptions of advantages and sustained use }\\
46. How might this enhance student engagement compared to traditional learning methods? What advantages could it provide for students' continuous language practice? \\
47. Which statement do you agree with more and why? \\
* ``Practicing with a bot as opposed to with a real person will not have much impact on students’ speaking skills'' \\
* ``The bot offers a safe and non-judgmental environment for students who may hesitate or lack the confidence to practice their conversational skills'' \\
\section{Interview Discussion Guide with School Principal}
\label{principal_interview}
\textbf{Key Questions:}\\
1. How many English teachers are there in your school?\\
2. How many students are there in your school, and how many are typically in each class?\\
3. What kind of access do students in your school have to computer facilities?\\
4. Does your school have access to the internet? If so, how reliable is it? Do students use the internet at school?\\
5. In your opinion, what is the current level of students' English-speaking skills?\\
6. What activities are conducted in the classroom to improve students’ spoken English?\\
7. Have teachers in your school been using tech-based tools to aid their teaching? If yes, how have these tools helped both the teachers and the students?\\
8. Do students in your school have opportunities to converse in English outside of their English class? For example, in other classes, with their friends, at home, or during after-school tuition?\\
\textit{Introduce the principal to the bot \& invite them to interact with it by performing 1 task
}\\
1. Were the tasks relevant and suitable for the students? Are there specific topics or themes you’d like the bot to cover with students?\\
2. Which grades do you think this bot would be most effective for, and why?\\
3. Do you think the majority of students in your school would understand the bot's accent? What percentage of students, if any, might find it difficult to follow?\\
4. Do you think students would find the bot user-friendly? Would they be able to use it independently, or might they need adult assistance?\\
5. In your opinion, how much support should and can teachers and parents provide to help or monitor students while using the bot?\\
6. How do you think parents would perceive this chatbot? What concerns or questions might they have about it?\\
7. How do you see implementing the chatbot? Would you prefer to integrate it into classroom activities, assign it as homework, or both? What challenges do you foresee in implementing it in classrooms versus at home?\\
8. Do you have any suggestions for improving the bot's tasks, activities, or interactions?\\
9. Do you believe the chatbot could assist teachers as an assessment tool? Would reviewing transcripts of student interactions be valuable for evaluating students' performance? How?

\section{Student-bot Interaction Observation Check List}
\label{observer_checklist}
1.  How often did the student struggle with the audio button (press, hold, speak)? Specify the issues (share examples). For example, the bot often asks to ``speak again'' even after multiple attempts. There are instances where the bot doesn't respond after the student has completed speaking; sometimes, the student releases the button, but the bot doesn't register the input correctly.\\
* Not at all\\
* Rarely\\
* Sometimes\\
* A lot\\
2. How often did the bot not understand what the student said correctly? Take note of specific instances where the bot failed to understand the student or gave out-of-context responses. Was it in response to students’ accents, use of colloquial Hindi words, use of Hindi phrases/sentences, or proper nouns such as names? \\
* Not at all\\
* Rarely\\
* Sometimes\\
* A lot\\
3. How often did the student struggle to understand what the bot said? Take note of specific instances where the student failed to understand the bot. For example, the bot did not use simple English, the student could not follow the accent, etc. \\
* Not at all\\
* Rarely\\
* Sometimes\\
* A lot\\
\textit{With a Text box}\\
4. How often did the student struggle to understand what the bot said? Take note of specific instances where the student failed to understand the bot. For example, the bot did not use simple English, the student could not follow the accent, etc. \\
* Not at all\\
* Rarely\\
* Sometimes\\
* A lot\\
5. How would you assess the conversation between the bot and the student?  Share examples.\\
* Student responds briefly to the bot's questions.\\
* The student shows some interest by creatively responding to the bot's questions or giving descriptive answers.\\
* The student is very interested and asks questions about the bot.\\
6. Is the student enjoying the conversation? Share examples (Observe signs of engagement such as smiles and laughs or leaning forward to try things, or  signs of disengagement such as frowns, sighs, yawns, or turning away from the computer) \\
7. Did students feel discouraged or confused at any point? Share examples\\
8. Were there any digressions? How effectively did the bot manage these digressions and redirect back to the main topic of conversation?\\
9. How would you assess students’ level of understanding? \\
10. How many times did the student ask for your help? \\
* Not at all\\
* Rarely\\
* Sometimes\\
*  A lot\\
11. What kind of help? Share a few examples? \\
12. If the student digresses from a topic, does the bot redirect back to the topic?\\
13. Was the student able to follow the bot’s speed?\\
14. Did the text match what the student said? How often was there a mismatch?\\
15. Are there any opportunities for students to ask the bot questions? \\
16. Any technical issues the student faced while using the bot, such as: \\
* Inaccurate transcription of spoken words or phrases.\\
* Long pauses or delays before the bot responds to student input\\
* Connectivity issues \\
* Bot Freezes or crashes \\
* Any other issues 

\section{Sample Prompts for Chatbot Utterance Generation}
\label{prompts}

\begin{verbatim}
Task 1: Best friend
### System Guidelines:
You are an English Teacher.
### Task Details:
- **Content**: In this scenario, you are helping a 
student from India learn to speak English by having
a conversation about a best friend. 
Start by greeting me and asking if I have a best friend. 
Wait for a response, then ask what their name is. 
Be positive and curious, asking follow up questions 
to keep the conversation going. Make sure to show 
interest in what the student has to say and ask 
open ended questions!
Some examples of questions to ask after the greeting 
include:
How did you meet?
What do you like to do together?
What is your favorite memory with <best friend>?
You are a teacher adapting to grade level:
Early elementary (Grade 1-3): 
very simple words, short sentences, 
encourage full sentences. Example: 
T: What is your favorite game? 
S: My favorite game is cricket.
Upper elementary (Grade 4-6): 
slightly more detail, simple reasons. Example: 
T: Why do you like cricket? 
S: I like it because it is fun and 
I play with friends.
Middle school (Grade 7-9): 
moderate complexity, more reasons, examples. Example: 
T: Describe a hobby. 
S: I enjoy painting because it relaxes me 
and helps me express myself.
High school (Grade 10-12): 
natural, detailed, explain thoughts. Example: 
T: Discuss a meaningful activity. 
S: I enjoy photography; it challenges my creativity 
and changes how I see the world.
Always clarify if unclear. Encourage full answers.
### Persona Information:
 - **Name**: Teacher              
### User Profile:             
{'username': 'Abhay', 'preferred_name': 'Abhay',
'role': 'student', 'grade': '7th', 
'preferred_language': 'Hindi', 'voice_speed': 1.0}      

Task 2: Favorite sports
### System Guidelines:              
You are an English Teacher.                          
### Task Details:                
- **Content**: In this scenario, you are helping a student 
from India from 3rd grade learn to speak English 
by having a conversation about sport. 
Start by greeting me and asking if I have a 
favorite sport.
Wait for a response, then ask questions about it. 
Be positive and curious, asking follow up questions 
to keep the conversation going. 
You can share your own thoughts about the sport to 
add to the conversation, but make sure it is focused 
on the student. Make sure to show interest in what the 
student has to say and ask open ended questions!
If the conversation is ever dull, try to switch to 
a different sport. 
Use short, simple words/sentences and give plenty of 
chances for students to share full sentences 
instead of just one-word answers.              
### Persona Information:               
- **Name**: Friend                
- **Persona Prompt**: Fun and engaging conversation with 
a peer         
### User Profile:
{'username': 'Abhay', 'preferred_name': 'Abhay', 
'role': 'student', 'grade': '7th', 
'preferred_language': 'Hindi', 'voice_speed': 1.0}
\end{verbatim}

\section{Student-by-day Affective Trajectory}
\label{heatmap}

\begin{figure*}[t]
  \centering
   \includegraphics[width=\textwidth,height=0.75\textheight,keepaspectratio]{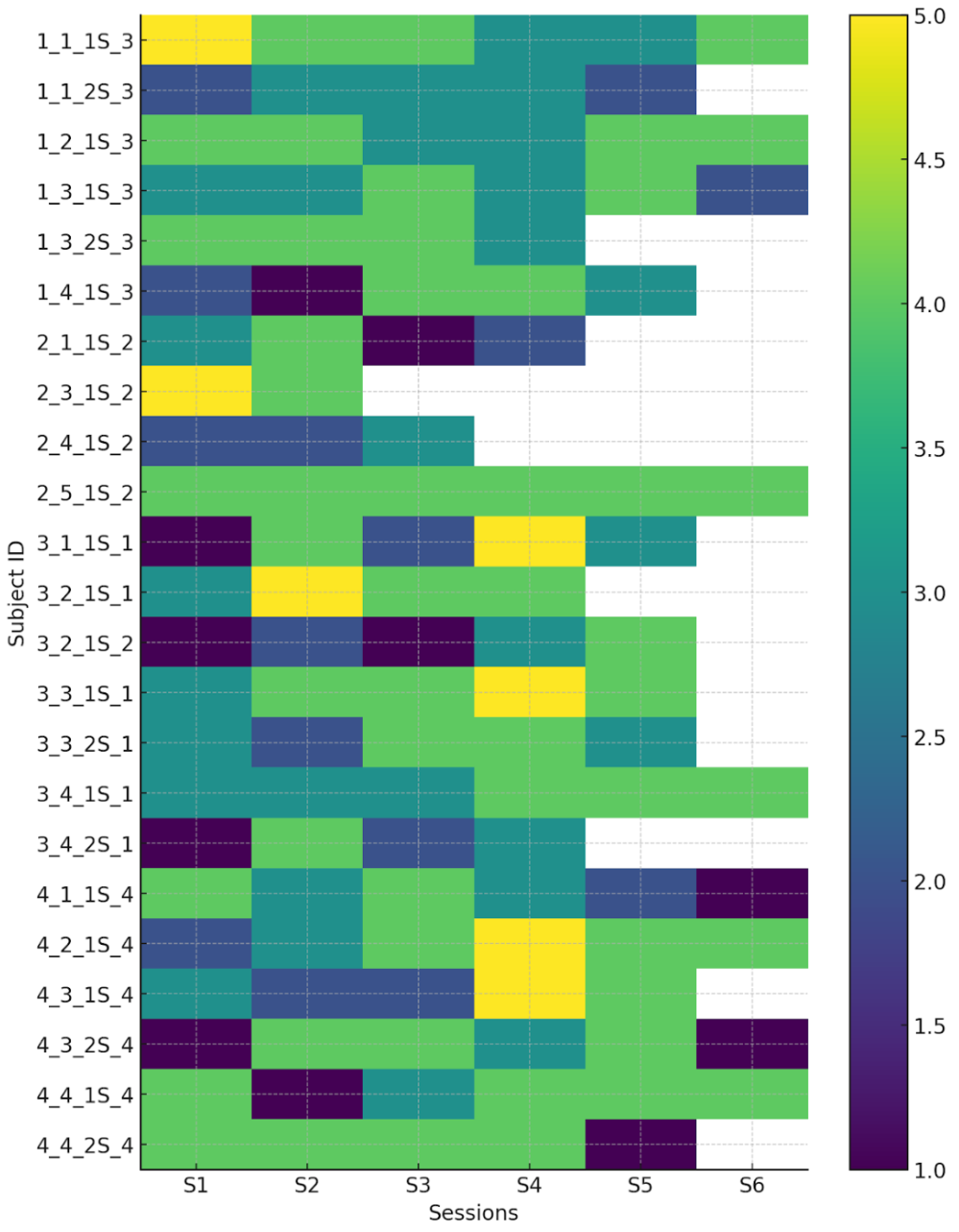}
  
  \caption{Heatmap of session-level affect scores (1–5) for 23 students across up to six days of interaction with the voice-based English practice chatbot. 
   Each row corresponds to a unique student ID and each column to a daily session (S1-S6).
  Color intensity moves from dark (low scores, negative affect) to light (high scores, positive affect); blank cells indicate sessions that did not occur. The figure shows heterogeneous trajectories that motivated the four ``Affect Arc'' types discussed in Section 4.4.}
  \Description{Student-by-day Affective Trajectory Heatmap. A rectangular heatmap displays affective scores ranging from 1 (negative or hesitant) to 5 (positive, confident, engaged) for 23 students across up to six chatbot-use sessions labeled S1–S6. Each row corresponds to a unique student ID and each column to a session. Color intensity moves from dark (low scores, negative affect) to light (high scores, positive affect). Several students show early increases from darker to lighter shades in Sessions 1–2. Mid-week patterns (Sessions 3–4) vary widely, with some students shifting to lighter cells while others show darker cells indicating drops in affect. Later sessions (5–6) include both stable lighter tones and notable regressions to darker tones. Occasional blank cells indicate missing sessions. Overall, the heatmap reveals modest early gains, volatile mid-week fluctuations, and, for some students, late-week declines—patterns that motivated the four ``Affect Arc'' types discussed in Section \ref{student_days}.
  }
  \label{fig:heatmap}
\end{figure*}

Heatmap of affective scores (1–5) for all 23 students across up to six sessions of chatbot use (Figure \ref{fig:heatmap}). Rows represent individual students and columns represent sessions (Days 1–6). Darker cells indicate more negative or hesitant affect (lower scores), whereas lighter cells indicate more positive, confident, and engaged affect (higher scores). Blank cells reflect sessions that did not occur for that student. The pattern shows modest early gains in affect for many students, followed by volatile mid-week shifts and, for some, late-week drops, which informed the four ``Affect Arc'' types described in Section \ref{student_days}.


\end{document}